# A space-efficient quantum computer simulator suitable for high-speed FPGA implementation

Michael P. Frank[a], Liviu Oniciuc[a], Uwe H. Meyer-Baese*[a], Irinel Chiorescu[b]
[a]FAMU-FSU College of Engineering, 2525 Pottsdamer St., Tallahassee, FL, USA 32310-6046;
[b]National High Magnetic Field Lab., 1800 E. Paul Dirac Dr., Tallahassee, FL, USA 32310-3706

## ABSTRACT

Conventional vector-based simulators for quantum computers are quite limited in the size of the quantum circuits they can handle, due to the worst-case exponential growth of even sparse representations of the full quantum state vector as a function of the number of quantum operations applied. However, this exponential-space requirement can be avoided by using general space-time tradeoffs long known to complexity theorists, which can be appropriately optimized for this particular problem in a way that also illustrates some interesting reformulations of quantum mechanics. In this paper, we describe the design and empirical space/time complexity measurements of a working software prototype of a quantum computer simulator that avoids excessive space requirements. Due to its space-efficiency, this design is well-suited to embedding in single-chip environments, permitting especially fast execution that avoids access latencies to main memory. We plan to prototype our design on a standard FPGA development board.

## 1. INTRODUCTION

Since a scalable, widely accessible quantum computer has not yet been built, it is important to be able to demonstrate the theoretical operation of quantum computers using simuators based on existing classical computing technology. Such tools are useful for the validation and testing of new quantum algorithms in research settings, as well as for the education of students as well as more experienced scholars who may be new to the emerging field of quantum computing.

Unfortunately, most or all of the existing widely-available quantum computer simulators are severely limited in the size of the quantum circuits that they can simulate. This is due to the fact that nearly all of the traditional simulators operate by updating an explicit representation of the quantum state vector of the simulated quantum circuit. In the worst case (which is also the case that is typically encountered in practice, in most of the interesting quantum algorithms), the number of nonzero elements of the state vector increases exponentially with the number of operations (gates) that are dynamically applied in the quantum circuit, and therefore increases exponentially with the size of the problem to be solved. This means that, even when a sparse representation of the state vector is used, the available memory on any given platform imposes a rather severe limit on the size of the quantum circuits that can be feasibly simulated.

For example, if a given machine has 8 GB of main memory, then it might only be able to simulate general quantum circuits containing 30 or fewer nontrivial gates, since representing the final quantum state of such a circuit would typically require storing $2^{30}$ = 1G eight-byte floating-point complex numbers. Furthermore, since accessing main memory (as opposed to on-chip caches) is relatively slow, the large amount of memory required for simulating even circuits of sizes somewhat below this limit can still impair the simulator's performance.

It would be desirable to have a simulator whose capabilities were not so strictly limited by the available memory, so that the simulator can be implemented on a fast single-chip hardware platform, and also so that it can (given sufficient time) simulate circuits of sizes beyond the limits of traditional simulators, with more graceful performance degradation.

Fortunately, computational complexity theorists have long been aware that there is a general algorithmic transformation that can be applied to reduce the space requirements of algorithms. The basic concept is simply to recalculate data values dynamically when needed, rather than storing them explicitly. As long as an algorithm's dataflow graph is not as deep as it is wide, using this approach can reduce the algorithm's space complexity.

* Corresponding author. umb@eng.fsu.edu; phone 1 850 410-6220; fax 1 850 410-6479; www.eng.fsu.edu

This condition applies to the simulation of quantum computers. Each element of the quantum state vector at a given step of a quantum algorithm typically depends only on the values of 1 or 2 elements of the state vector at the preceding step. These values can be recomputed on demand from the values at the next preceding step, and so forth, in recursive fashion. The recursion back through the entire history of $N$ previously applied quantum operations requires only $O(N)$ space on a stack, to keep track of the result accumulated so far at each node of the current path through the dataflow graph.

Possibly the first person to realize that this general kind of procedure could be applied to the calculation of quantum-mechanical amplitudes was the famous physicist Richard Feynman, who in his dissertation work[1] showed how quantum mechanics could be reformulated in terms of a quantity he called the *path integral*, which essentially amounted to a continuous analogue of a sum over paths through a discrete dataflow graph.

When the complexity theory of quantum computing was being developed in the early 1990s, it was quickly realized[2] that the same idea, back now in the discrete realm, could be applied to the simulation of quantum computers as well, leading to the important complexity-theoretic relation that BQP $\subseteq$ PSPACE, where BQP is the set of problems solvable by probabilistic quantum algorithms with a polynomial number of operations (as a function of problem size), and PSPACE is the set of problems solvable by classical computers using a polynomial amount of memory. More generally, we can show[3] that a quantum algorithm with $s$ qubits and $t$ operations can be simulated using space $O(s + t)$.

Although this essential insight has been known for at least 16 years now, to our knowledge it has not yet been applied to develop a flexible and widely-available tool for simulating quantum algorithms in such a way that the available memory is not a significant limiting factor on the size of the quantum computations that can be simulated. It is the goal of the SEQCSim (say "SEEK-sim") project at Florida State University to remedy this situation by providing flexible, well-optimized freely-available software and hardware implementations of a <u>S</u>pace-<u>E</u>fficient <u>Q</u>uantum <u>C</u>omputer <u>Sim</u>ulator.

So far, we have developed a working software prototype of our simulator in C++, and have empirically demonstrated its correctness and space-efficiency on a variety of simple test cases. We present some of these results in sections 2-4. Next steps include the development of a more powerful programming environment for the software version of the simulator, as well as a performance-optimized special-purpose hardware implementation of the simulator, to be prototyped using a standard FPGA (field-programmable gate array) platform, which we will describe in section 5. Section 6 concludes.

## 2. SEQCSIM ALGORITHM

The presently-available software prototype of our simulator (version 0.8) operates according to a simple procedure.

The simulator first loads a definition of the quantum circuit to simulate from a set of four ASCII text input files, called `qconfig.txt`, `qinput.txt`, `qoperators.txt`, and `qopseq.txt`, examples of which are shown in listings 2-5. These files are structured in a simple throwaway file format, which will likely be replaced in later versions of our simulator by a high-level quantum programming language based on C++. The configuration file, given in `qconfig.txt`, specifies the width of the quantum circuit, and assigns various named registers to specific bit-fields within it. The quantum algorithm, specified explicitly as a gate sequence in `qopseq.txt`, may use any fixed-width quantum gates, whose matrix elements are given in `qoperators.txt`. The initial input state (which must be a classical state in the computational basis) is given in `qinput.txt`.

The general approach of the simulator is to progress forwards through the quantum algorithm (circuit) one operation (gate) at a time, while keeping track of the amplitude of only a single basis state, in the classical computational basis, which is selected pseudo-randomly at each step in accordance with the flow of probability mass in the quantum algorithm, in such a way that the simulator's probability of ending up at each final basis state precisely matches what would be obtained from a complete calulation of the final quantum state vector.

This approach evokes an old interpretation of quantum mechanics by Bohm[4-5], who showed that a quantum system can be conceived of as having a unique classical state at each time which evolves (either deterministically or nondeterministically) in accordance with the probability current through the system's phase space that is induced by the Schrödinger time-evolution. In this model, a complete wavefunction still exists mathematically, but it is conceived of as just being a "pilot" wave that guides the evolution of the physical state, rather than being thought of as being the actual physical state itself. In accordance with the subordinate status of the wavefunction in Bohm's philosophy, rather than storing the entire wavefunction, and conceiving of it as being the simulated state, we only calculate values of the wavefunction at points that are needed to compute the transition probabilities along the specific possible trajectory through the classical configuration space that is presently being explored.

The core algorithms for updating the stochasically-evolved basis state (procedure "run()") and calculating wavefunction amplitudes for specific basis states (function "calcAmp()") are outlined below in Listing 1.

To see a detailed graphical illustration of the functioning of this algorithm on a particularly simple example circuit, please refer to our previous paper[3].

Listing 1. Outline of the core algorithm used in the present version of SEQCSim. The selection of a particular operation (gate and operand bits) on lines 5 and 18 determines a set of possible "neighbor" or "predecessor" basis states of the current one, differing from the present state on the operand bits.

```
1   procedure SEQCSim::run():
2       curState := inputState;           // Current basis state, in the computational basis
3       curAmp := 1;                      // Amplitude of current basis state
4       for PC =: 0 to #gates,            // Index of current operation in the gate sequence
5           with respect to the operator gate[PC] and its operands,
6               for each neighbor nbr_i of curState,
7                   if nbr_i = curState, amp[nbr_i] :=curAmp;
8                   else amp[nbr_i] := calcAmp(nbr_i);
9               amp[] := opMatrix * amp[];              // Complex matrix product
10      prob[] := normSqr(amp[]);         // Calc probs as normalized squares of amplitudes.
11      i := pickFromDist(prob[]);        // Pick a random successor of the current state.
12      curState := nbr_i;                // Go to that neighbor.
13      curAmp := amp[nbr_i].             // Remember its amplitude, calculated earlier.
14
15  function SEQCSim::calcAmp(State nbr):     // Recursive amplitude-calculation procedure
16      curState := nbr;
17      if PC=0, return (curState = inputState) ? 1 : 0;  // At t=0, input state has all the amplitude.
18      else, with respect to the operator gate[PC−1] and its operands,
19          for each predecessor pred_i of curState,
20              PC := PC − 1;
21              amp[pred_i] = calcAmp(pred_i);     // Recursive calculation of pred. amp.
22              PC := PC + 1;
23          amp[] := opMatrix * amp[];           // Complex matrix product
24      return amp[curState];
```

## 3. EXAMPLE QUANTUM CIRCUIT USED IN TESTING

For purposes of testing the correctness and performance of our algorithm, we focused on a simple family of in-place binary adder circuits based on an algorithm by Draper[6]. These adders use a Quantum Fourier Transform (QFT) and its inverse to convert one of the addends into and out of a phase representation, and uses phase gates between addends to carry out the addition in the phase representation, thereby avoiding the need to explicitly compute carries. These adders are not particularly efficient (since they require order $n^2$ gates for an $n$-bit add) but they require no ancilla bits, and they provide a good test case that includes both trivial and nontrivial gates, and that spreads out and later reconcentrates amplitude in an interesting way. For our purposes, a "trivial" gate means a gate, like the phase gate described below, whose unitary matrix is diagonalizable in the computational basis. With such gates, each basis state has only one possible predecessor, so these gates do not have a very significant impact on the time complexity of the simulation – in our approach, any sequence of trivial gates is simulated in linear time.

An example of the adders used is shown in fig. 1 below. This illustration was prepared using the freely-available QCAD design/simulation tool, version 1.96, available from `http://apollon.cc.u-tokyo.ac.jp/~watanabe/qcad/index.html`. In this figure, H represents the Hadamard gate $H = (\sigma_x + \sigma_z)\cdot 2^{-1/2} = [1, 1; 1, -1]/2^{1/2}$, or in displayed form,

$$H = \frac{1}{\sqrt{2}}\begin{bmatrix} 1 & 1 \\ 1 & -1 \end{bmatrix}. \tag{1}$$

A number $k$ in a box represents a controlled-phase gate $\varphi_k$ for a phase rotation of $k$ degrees. In terms of the rank-2 identity operator $\hat{I} = [1, 0; 0, 1]$, number operator $\hat{n} = [0, 0; 0, 1]$, its complement $\overline{n} = \hat{I} - \hat{n}$, and tensor product $\otimes$, and then also written out more explicitly as a displayed matrix, this operator can be defined as

$$\begin{aligned}\varphi_k &= \overline{n} \otimes \hat{I} + \hat{n} \otimes \exp(\mathrm{i}\hat{n}k\pi/180) \\ &= \begin{bmatrix} 1 & 0 & 0 & 0 \\ 0 & 1 & 0 & 0 \\ 0 & 0 & 1 & 0 \\ 0 & 0 & 0 & \exp(\mathrm{i}\pi k/180) \end{bmatrix}. \end{aligned} \tag{2}$$

The phase gate rotates the phase of a given basis state by the specified number of degrees if and only if both of the input bits are 1. It is symmetrical with respect to the control and target bits. It is a *trivial* gate – note its matrix is diagonal.

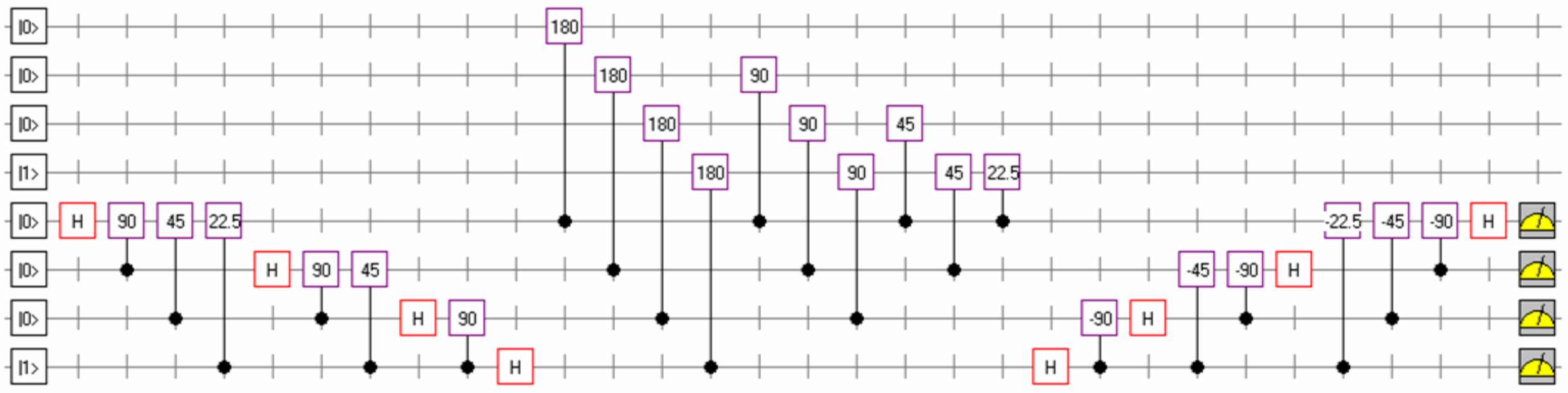

Fig. 1. Illustration, using the freely-available QCAD tool, of a quantum circuit for adding two 4-bit binary numbers $a$, $b$ in place using Draper's algorithm. The top group of 4 qubits represents $b$, the bottom four qubits are $a$, and the most-significant qubit in each group is at the top. The initial state shown at the left is $a$=1, $b$=1. The first (leftmost) 10 gates perform a quantum Fourier transform (QFT) of $a$ in-place, to convert the value of $a$ into a pattern of phases on the amplitudes over the $a$ subspace. The next 10 gates increment the phases by the value of $b$. The final 10 gates perform an inverse QFT to convert the phases back into a value of $a$. The overall operation performed is $a := a + b$, and the final value of $a$ (which is measured after the computation) is 2. The value of $b$ is unchanged.

The above example circuit can be easily prepared for input into the SEQCsim simulator by describing it in a simple text input format in the four files {`qconfig`, `qinput`, `qoperators`, `qopseq`}`.txt`, as illustrated in listings 2-5 below. The precise format of these files has some limited flexibility – keywords may be abbreviated, whitespace is ignored, and lines beginning with "comment:" are ignored. The format specifier on the first line allows for future extensions of the file format, while allowing new versions of the simulator to remain backwards-compatible with older input files.

Listing 2. Contents of the ASCII text input file `qconfig.txt`, which is used to tell SEQCsim the size and registers of the input circuit, for the circuit shown in fig. 1. (Line numbers shown at the left are not included in the file.)

```
qconfig.txt format version 1
bits: 8
named bitarray: a[4] @ 0
named bitarray: b[4] @ 4
```

Listing 3. Contents of the ASCII text input file `qinput.txt` used to tell SEQCsim the decimal values of the input registers for the circuit shown in fig. 1. (Line numbers shown at the left are not included in the file.)

Listing 4. Contents of the ASCII text input file `qoperators.txt` used to tell SEQCsim the definitions of the quantum operators (gates) used in the circuit shown in fig. 1. (Line numbers at the left are not included in the file.)

```
qoperators.txt format version 1
operators: 8
operator #: 0
name: H
size: 1 bits
matrix:
(0.7071067812 + i*0)(0.7071067812 + i*0)
(0.7071067812 + i*0)(-0.7071067812 + i*0)
operator #: 1
name: cPiOver2
size: 2 bits
matrix:
(1 + i*0) (0 + i*0) (0 + i*0) (0 + i*0)
(0 + i*0) (1 + i*0) (0 + i*0) (0 + i*0)
(0 + i*0) (0 + i*0) (1 + i*0) (0 + i*0)
(0 + i*0) (0 + i*0) (0 + i*0) (0 + i*1)
```

***... (six additional operators elided for brevity)…***

Listing 5. Contents of the ASCII text input file `qopseq.txt` used to tell SEQCsim the sequence of quantum operations (gate instances) used in the circuit shown in fig. 1. (Line numbers shown at the left are not included in the file.)

```
qopseq.txt format version 1
operations: 30
operation #0: apply unary operator H to bits a[3]
operation #1: apply binary operator cPiOver2 to bits a[3], a[2]
operation #2: apply binary operator cPiOver4 to bits a[3], a[1]
operation #3: apply binary operator cPiOver8 to bits a[3], a[0]
```

***...(22 additional gate operations elided for brevity)…***

```
operation #26: apply binary operator inv_cPiOver8 to bits a[3], a[0]
operation #27: apply binary operator inv_cPiOver4 to bits a[3], a[1]
operation #28: apply binary operator inv_cPiOver2 to bits a[3], a[2]
operation #29: apply unary operator H to bits a[3]
```

Listing 6. Text output from SEQCSim when run with the above text files (listings 2-5) as input. (Line numbers shown at the left are not included in the output.) Note that at the conclusion of the computation, the value of register *a* (least significant 4 bits) is $0010_2 = 2$, verifying that the simulator has correctly determined that 1 + 1 = 2.

```
Welcome to SEQCSim, the Space-Efficient Quantum Computer SIMulator.
    (C++ console version)
By Michael P. Frank, Uwe Meyer-Baese, Irinel Chiorescu, and Liviu Oniciuc.
Copyright (C) 2008-2009 Florida State University Board of Trustees.
    All rights reserved.
```

***..(2 blank lines)..***

```
SEQCSim::run(): Initial state is 7->00010001<-0 (8 bits) ==> (1 + i*0).
SEQCSim::Bohm_step_forwards(): (tPC=0)
   The new current state is 7->00011001<-0 (8 bits) ==> (0.707107 + i*0).
SEQCSim::Bohm_step_forwards(): (tPC=1)
   The new current state is 7->00011001<-0 (8 bits) ==> (0.707107 + i*0).
```

***…(26 intermediate steps elided for brevity)…***

```
SEQCSim::Bohm_step_forwards(): (tPC=28)
    The new current state is 7->00011010<-0 (8 bits) ==> (0.707107 + i*0).
SEQCSim::Bohm_step_forwards(): (tPC=29)
    The new current state is 7->00010010<-0 (8 bits) ==> (1 + i*0).
SEQCSim::done(): The PC value 30 is >= the number of operations 30.
    We are done!
```

## 4. EMPIRICAL MEASUREMENTS OF SPACE/TIME COMPLEXITY

To show that the space and time complexity of SEQCSim indeed respond in the expected manner to changes in the size of the simulated circuit, we used the following procedure. The second author wrote a tool in C# to automatically generate the required `qconfig.txt` and `qopseq.txt` input files for SEQCsim, as well as corresponding `.qcd` circuit files for QCAD, for Draper adders of any desired number and size of operands. Using this tool, we generated adders of size 2×2 (2 addends, 2 bits each) up through size 2×14 (2 addends, 14 bits each), and ran QCAD and SEQCsim on each one, on a typical Dell desktop running Windows Vista, while measuring each application's peak memory usage and CPU time using the shareware Kiwi application monitor. The results were tabulated and used to generate the charts below.

Fig. 2 shows that overall memory usage of QCAD (which presumably internally uses a traditional state-vector based simulation technique) asymptotically increases exponentially with the circuit size, whereas the memory usage of SEQCsim remains essentially flat – the memory used internally by our algorithm is only in the kilobytes, and in practice is always dominated by the space required for the standard C++ libraries which we invoke to provide functions such as text I/O and pseudo-random-number generation.

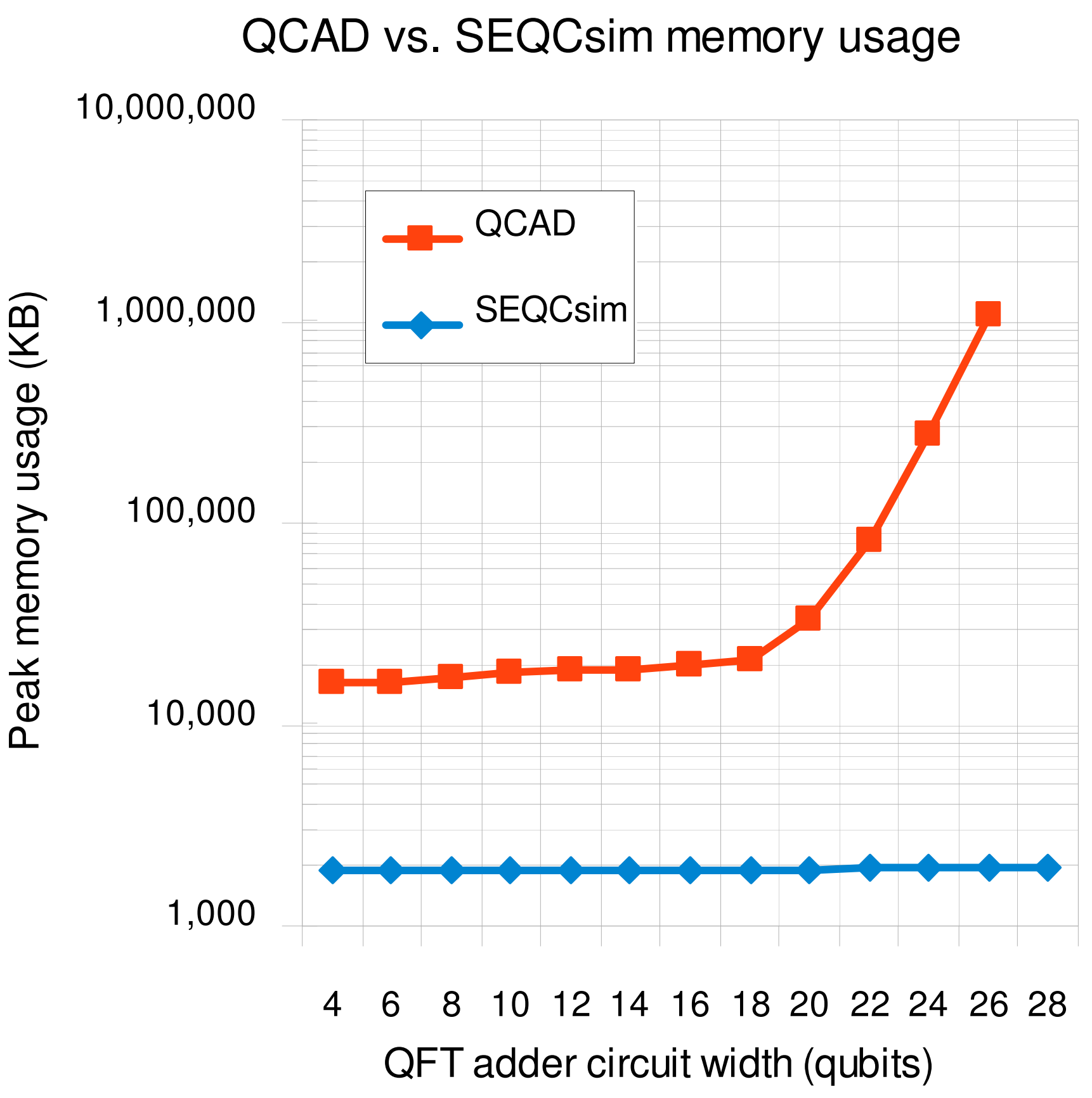

Fig. 2. Overall peak memory usage, in kilobytes, of QCAD versus SEQCsim, for Draper adder circuits of width 4 (2×2) through 28 (2×14), as measured using the Kiwi application monitor. These figures include pages allocated for sharable DLLs, but a comparison of private working set sizes, as measured by the Windows Task Manager in Vista, gives qualitatively similar results. QCAD's higher base memory usage is unsurprising, since it requires more libraries to support its GUI. Note that the vertical scale is logarithmic. Beyond about 18 bits, QCAD's internal memory usage can be seen to be increasing exponentially, as the size of its dynamic data set exceeds the memory requirements of its base API libraries; this behavior would be expected for any simulator based on an explicit state-vector representation. Note that, in comparison, SEQCSim's memory usage remains essentially flat, at about 2 MB, throughout this range, and most of this is attributable to the system & language libraries used. Note there is no data point for QCAD for circuit width 28, because the required memory (about 4 GB) exceeded what was available on the PC that was used for testing.

The next chart, shown in fig. 3, shows in more detail how the memory usage of SEQCsim increases as a function of the number of gates in the quantum circuit (for the same runs as fig. 2). There are some irregularities in the graph, which we hypothesize result from the fact that the number of pages allocated by the dynamic memory allocator may vary slightly from run to run depending on unpredictable factors. These irregularities could be minimized by averaging the results over multiple runs, but we have not yet done this. Despite the irregularities, there is a clear almost-linear trend in the memory usage which is well accounted for by the increasing number of stack frames that must be allocated as the number of levels of calcAmp() recursion (which is proportional to the depth of the circuit) increases. The size of these stack frames is somewhat larger than needed, because our current implementation is somewhat inefficient in that it stores the values of a variety of temporary lexical variables at each level of the recursion. It would be possible to eliminate this inefficiency through a more direct implementation that replaced the recursion with a more specialized iterative routine for traversing the tree of predecessors that reused its temporary variables. This routine would only need to keep track of a single complex number (the accumulated amplitude) at each level of the tree, *i.e.* for each gate.

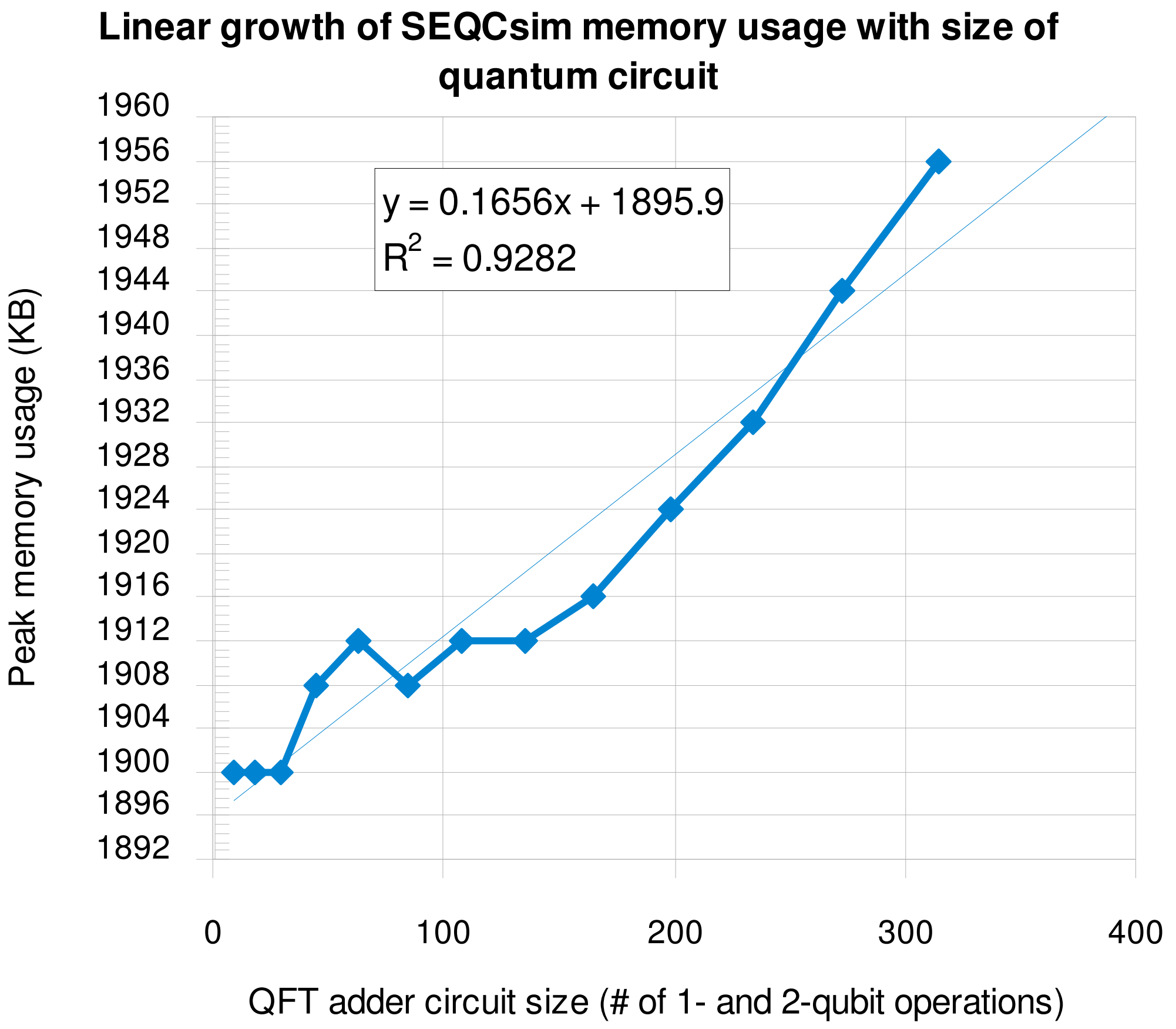

Fig. 3. This graph shows how the peak memory usage of SEQCsim varied with the size in gates of the QFT-based adder circuits that were used for testing. The vertical scale here is linear, and we can see there is a slight, roughly linear increase in memory usage from 1900 KB to 1956 KB, in page-sized 4K increments, as we go from 9 operations for the 2×2 adder to 315 operations for the 2×14 adder. The results are somewhat noisy due to slight runtime variations in pages allocated by the memory manager because these results were not averaged over multiple runs. The extra 56K required for the largest circuit size is easily accounted for by the increased number of stack frames that must be allocated in order to get through all 315 levels of recursion in the calcAmp() function in the final step. If desired, the size of these stack frames could be further reduced from the current maximum of ~187 bytes to only a single complex number (say about 8 bytes), by replacing the recursive function with a more specialized iterative tree traverser.

Finally, fig. 4 shows how the CPU time used by both QCAD and SEQCsim varies with the width of the simulated circuit. At present, SEQCSim is about a factor of 2 slower than QCAD for larger circuit sizes. This is unsurprising, given that presently SEQCSim is highly flexible (gates of any width can be defined) and uses a relatively elaborate recursive procedure that invokes multiple levels of C++ abstraction, as opposed to the more straightforward state-vector updating that must be done in QCAD. There is much room for further improvement in SEQCSim's performance.

It would be fairly straightforward to modify SEQCSim to carry out an ordinary state-vector representation of the state until the limit of memory (or on-chip cache) is reached, and then revert to the recursive amplitude-calculation procedure only for further state evolution beyond that point. This would allow us to take full advantage of the available memory to boost performance, while retaining the ability to handle circuits of larger sizes without crashing.

Alternatively, we can keep SEQCsim's memory usage minimal, while giving it a substantial constant-factor performance boost by reimplementing its kernel using a custom or semi-custom hardware architecture. That approach will be discussed in the next section.

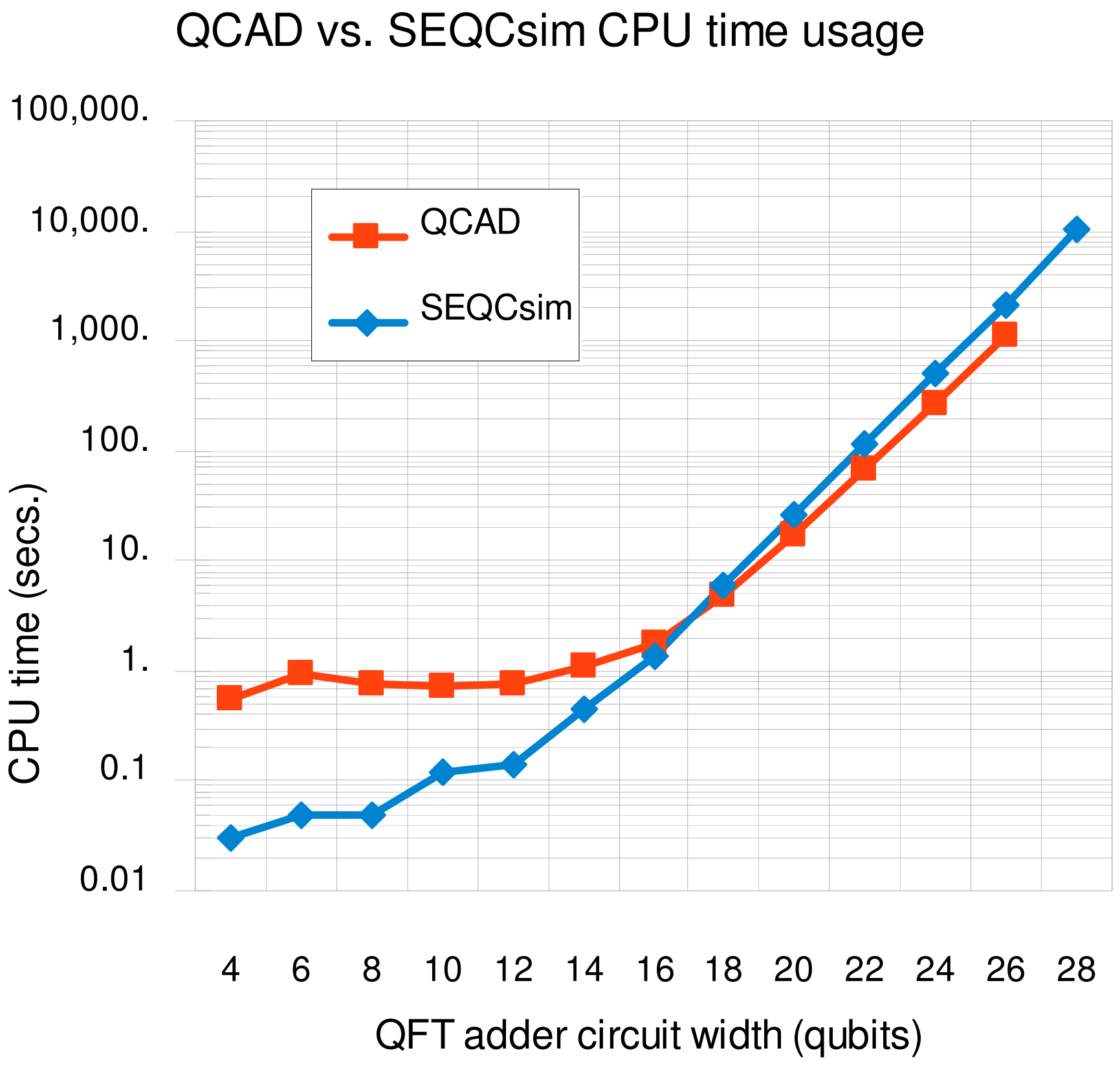

Fig. 4. Comparison of CPU time used by QCAD vs. SEQCSim for 2×2 through 2×14 bit adders. QCAD's greater CPU time usage for small circuit sizes can be accounted for by the fact that it has a GUI which is used to load the circuit and display results, whereas SEQCSim presently does not. We can see that the CPU time for both algorithms asymptotically increases exponentially, as expected, as the circuit width increases. Each increase in circuit width by 2 bits results in about a factor of 4 increase in time complexity, as expected. In the case of QCAD, this is because the state vector is 4× larger; whereas for SEQCSim it is because, for each additional bit in addend $a$, there are 2 additional non-trivial gates (Hadamard gates) in the circuit. The performance of SEQCSim could be significantly improved (beating that of QCAD) by either leveraging additional memory (recalculating fewer amplitudes), or by reimplementation in special-purpose, single-chip hardware, an approach which is made feasible because of SEQCSim's low memory usage.

# 5. FPGA-BASED EMBEDDED ARCHITECTURE

Since the SEQCSim simulator needs little memory but has intensive arithmetic requirements, an FPGA-based hardware accelerator is currently being designed to improve simulator performance. Speed-up factors of 10-100× are typically achieved with FPGA-based accelerators[7]. In this section we discuss some design approaches we are exploring.

## 5.1 Hardware Resources and Design Tools in FPGA Environments

FPGAs have a large field of arithmetic resources that can be tailored to the algorithms needed. The new generation of FPGAs like Xilinx Virtex have over 500 embedded 18×18-bit multipliers, over a hundred 18 Kb memory blocks, and over 100,000 logic blocks that can be used, for instance, as 3000 32-bit adders. In contrast, today's cell-based ASIC designs are relatively expensive (with mask charges of $4M in 60 nm technology) and are often replaced by FPGA-based solutions. The FPGA market share is growing fast; the two leaders (Altera and Xilinx) report revenues of over $1B annually. To make efficient use of the large arithmetic resource available in FPGAs, two system level design approaches are currently being considered.

The first is based on the use of Altera's new C2H compiler[8] which runs on any NIOS 2 based system. NIOS 2 is a royalty-free 32-bit soft-core microprocessor that can be configured with 1-5 pipeline stages, with or without data and program caches, and many peripherals like UART or SDRAM interfaces. The C2H compiler allows a quasi-automatic conversion of ANSI-C code into FPGA hardware. Simply mark the function in your C code you would like to accelerate (fig. 5) and the C2H compiler translates the C code into hardware, including register, arithmetic, memory blocks and the required Avalon Bus interface. These new blocks are added into the SOPC builder files that also allow running testbenches in a gate level simulator. Additional C coding techniques and compiler directives can be used in the software coding of the application to force the use of on-chip or off-chip memory, embedded multipliers or LUT-based multipliers. The features of C2H can be summarized as follows[8]:

- Tight integration with software design flow
- Push-button acceleration of ANSI/ISO C code
- Direct connection of hardware accelerators to CPU's memory map
- Seamless support for pointers and arrays
- Efficient latency-aware scheduling and pipelining of memory transactions

Since C2H provides automatic parallelization, the speed-ups are more substantial than the previous often-used custom instruction interface to NIOS, that can only use a register bank to exchange data between the host processor and co-processor. C2H user applications obtain[8] performance improvements exemplified by a convolution encoder (13×), FFT (15×), and matrix rotation (73×). In contrast, a 256-point Nios FFT custom user function has been reported with 45%-77%, *i.e.* less than a factor of 2 improvement[7, p. 632].

A second approach that is becoming more popular is the use of an application-specific microprocessor with a custom instruction set that can be tailored to the problem at hand avoiding the HW/SW partitioning bottleneck. Mixed architecture description languages (ADLs) like EXPRESSION, HMDES, or LISA that combine both the structural and behavioral details of the microprocessor architecture are preferred[9]. The language for instruction set architecture (LISA), for instance, allows us to specify a processor instruction or cycle accurately using a few LISA *operations*, then to explore the architecture using a tool generator (for the assembler, linker, and C-compiler) and profiler (fig. 6), and finally determining the speed/size/power parameters via automatically synthesized HDL code[10,11]. Processor models like 32-bit 5 pipeline stage LT-RISC, 32-bit 3-pipeline stage LT-DSP, or LT-VLIW 5-pipeline stage are provided and can easily extended for the custom QC instructions.

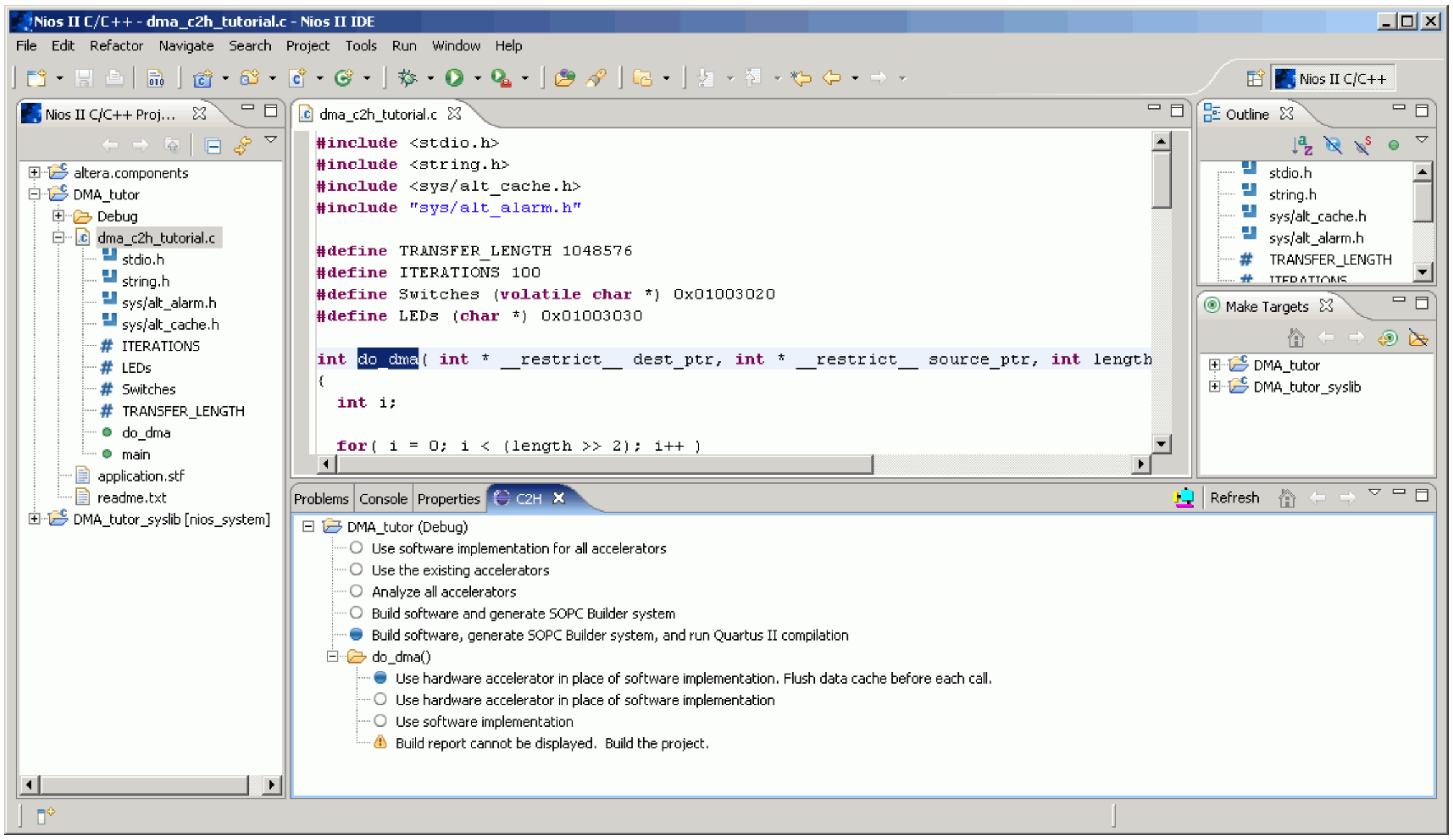

## 5.2 Design concepts for an FPGA-based SEQCSim

In the present software-only implementation of SEQCSim in C++, the execution time of the compute-intensive kernel in the recursive calcAmp() pseudocode is dominated by operations such as basic bit manipulation (extraction and modification of individual bits and small groups of bits in packed bit-vectors representing classical basis states of the quantum computer), basic arithmetic (addition and multiplication of floating-point complex numbers representing state amplitudes), and control-flow operations implementing loops, conditionals, and recursive procedure calls.

A custom multi-ported register structure for holding the current basis state being explored would be useful for fast extraction and modification of operand bits.

Fig. 5. Software speed-up development with C2H. User places instruction to be accelerated in a separate function. To accelerate this function, right click on it and select “Accelerate with the Nios II C2H Compiler”. Then run auto update of the whole system through compilation, *i.e.*, build software, add new units in SOPC builder, and Quartus compilation produces a configuration bit stream that is downloaded to the FPGA. Compile time approximately 1 hour.

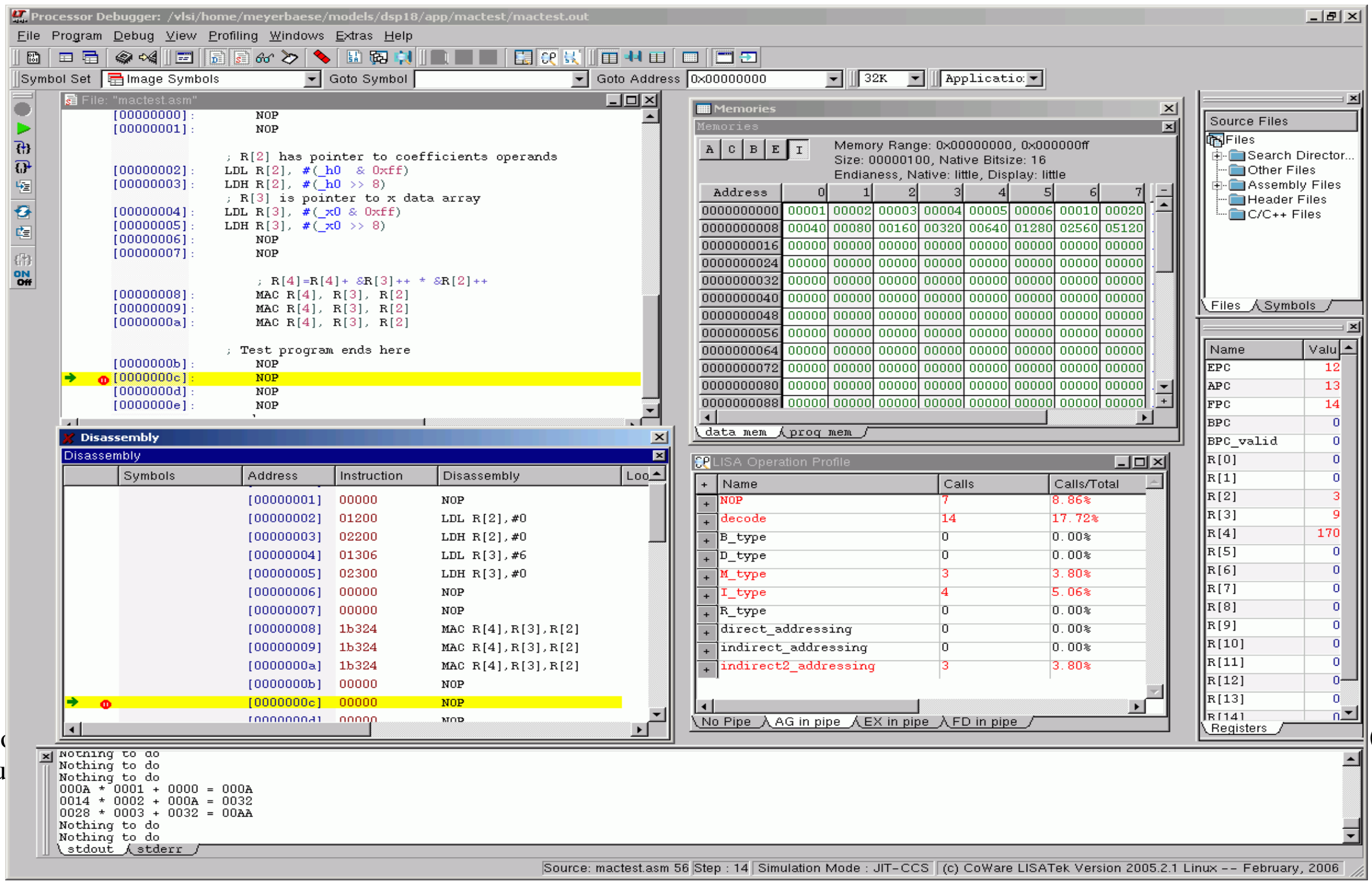

Fig. 6. Screenshot of LISA development tools: Disassembler, memory monitor, pipeline profiles, files and register window.

A specialized multiply-accumulate unit for complex numbers based on a custom floating-point number representation (which need not conform to the IEEE 754 standard) can significantly speed up calculation of complex amplitudes.

The recursive C++ routine implementing calcAmp() can be rewritten as a compact iterative kernel in plain C, which can then be automatically transformed into equivalent dataflow & state-machine hardware by a tool such as Altera's C2H compiler, thereby eliminating much of the control-flow overhead.

A custom stack memory structure can be designed to replace the role of the C procedure-call stack. Instead of storing a full C++ stack frame, it can store just the minimal information needed by the kernel, mainly just a complex number tracking the amplitude accumulated so far at a given level of the tree of possible trajectories (*i.e.* for a given gate).

Finally, the available RAM on the FPGA chip can be configured as an associative cache memory to store recently-calculated amplitudes with LRU replacement, which will significantly speed up the recursive amplitude calculation, as it will avoid redundant recalculation of amplitude values insofar as possible, given the limit of available memory.

Using techniques such as the above, we estimate that we can obtain a speedup of at least about 50× in our FPGA-based implementation of SEQCSim, as compared with our present C++ software prototype running on standard PC hardware, at which point SEQCSim will be significantly faster than existing simulators, as well as able to handle larger circuits.

## 6. CONCLUSION AND FUTURE WORK

We have developed and demonstrated a working software prototype of an extremely memory-efficient quantum computer simulator, which will soon be released publicly through the site http://www.eng.fsu.edu/~mpf/SEQCSim.htm. Our prototype uses an amount of memory that increases only linearly in the size of the quantum circuit being simulated, at a proportion which was determined to be less than two hundred bytes of memory per gate in the present study.

A more carefully-optimized implementation should be able to do even better then this, and achieve an asymptotic (*i.e.* marginal, amortized) memory usage of only 1 bit per qubit in the simulated circuit, plus 1 complex number per typical nontrivial gate (such as a Hadamard gate or a general controlled-$U_2$ gate).

Due to its miniscule memory requirements, a simulator of this class would be quite amenable for implementation in custom or semi-custom special-purpose hardware architecture, which can be easily prototyped using FPGAs loaded with embedded soft-core microprocessors such as Altera's NIOS or Xilinx's MicroBlaze, and/or a semi-custom processor designed with the help of tools such as Altera's C2H or the LISA tool set. Such a hardware-augmented implementation is expected to be able to outperform traditional software-only quantum computer simulators by a factor of 50-100×.

As for the software-only version of our simulator, its programming interface is presently rather cumbersome, requiring the user to define his or her quantum gates and gate sequences explicitly using an ad-hoc text input format. However, it would be straightforward to reimplement the simulator as, say, a set of classes in C++, or other object-oriented language, which would allow the programmer to describe quantum algorithms using the full expressive power of the host language, while observing a statistical behavior of his Qubit objects that matches what would be obtained from an ideal quantum computer, whose operation can be mimicked by a version of our simulator that is running "behind the scenes."

This future version of our simulator would be a near-ideal tool for allowing students and researchers to venture into quantum programming and freely experiment with new (and old) quantum algorithms without having to either learn a new programming language first, or worry about running out of memory.

## ACKNOWLEDGEMENTS

The authors gratefully acknowledge support for this work from the Council on Research and Creativity (CRC) at Florida State University. Irinel Chiorescu would also like to thank the National Science Foundation and the Defense Advance Research Projects Agency for his support under grants DMR-0645408 and HR0011-07-1-0031 respectively. The authors would also like to thank Altera Inc. and CoWare Inc. for their support under the University programs. Any opinions, findings, and conclusions or recommendations expressed in this paper are those of the authors, and do not necessarily reflect the views of the sponsors.